\documentclass[aps,amsmath,showpacs,twocolumn]{revtex4}
\usepackage{graphicx}

\begin{document}


\title{Natural Dark Energy}

\author{Douglas Scott} \email{docslugtoast@phas.ubc.ca}
\author{Ali Frolop} \email{afrolop@phas.ubc.ca}
\affiliation{Department of Physics \& Astronomy,
University of British Columbia, \\
Vancouver, BC, V6T 1Z1  Canada}

\date{1st April 2007}

\begin{abstract}
It is now well accepted that both Dark Matter and Dark Energy are required
in any successful cosmological model.  Although there is ample evidence that
both Dark components are necessary, the conventional theories make no
prediction for the contributions from each of them.  Moreover, there is
usually no intrinsic relationship between the two components, and no
understanding of the nature of the mysteries of the Dark Sector.
Here we suggest that if the Dark Side is so seductive then we should not be
restricted to just 2 components.  We further suggest that the most natural
model has 5 distinct forms of Dark Energy in addition to the usual Dark
Matter, each contributing precisely equally to the cosmic energy density
budget.
\end{abstract}


\maketitle

\date{today}

\noindent
The idea that most of the matter in the Universe is in a cold, dark, nearly
collisionless form has become firmly enshrined as part of the Standard Model
of Cosmology.  Like the expansion of the Universe in the 1940s or the Big
Bang paradigm in the 1970s, the existence of Dark Matter (DM) is now only
discounted by a handful of curmudgeons and crackpots \cite{legal}.
Part of the reason why DM has become so old-hat is that its enigmatic nature
has been eclipsed by an even weirder flavour of darkness.

Perhaps the greatest mystery in the whole of modern science is
the nature of the so-called Dark Energy (DE) \cite{DE}.
This bizarre component
drives the accelerated expansion of the Universe and dominates the overall
energy budget, making the geometry of space very close to flat.
The most prosaic explanation for the DE is that it is simply the energy
density of the vacuum, and we are then left simply with the problem of why
its obvious value (one Planck mass per Planck volume) would be about
$10^{120}$ times bigger than the measurements indicate \cite{why120}.

There are many theoretical ideas for the physical basis of the DE.  However,
it is fair to say that none of them are particularly well motivated, and
many of them appear like acts of desperation on the part of
theorists \cite{crazy}.

The evidence for DE comes from measurements like those of distant supernovae,
with the effects of the DE coming through the
evolution of the background cosmology via appropriate integrals over the
reciprocal of the Hubble expansion rate.
The Friedmann equation can be written in terms of the Hubble parameter as
\begin{align}
H(z) = H_0\big\{ & \Omega_{\rm R}(1+z)^4 + \Omega_{\rm M}(1+z)^3
 \notag\\
  + &\Omega_{\rm K}(1+z)^2 + \phantom{\Omega_{\Lambda}(1+z)^1}
 \! + \Omega_{\Lambda}(1+z)^0\big\}^{1/2},\notag
\end{align}
supplemented with the constraint equation
\begin{equation*}
\Omega_{\rm R} + \Omega_{\rm M} + \Omega_{\rm K} + \Omega_{\Lambda} = 1.
\end{equation*}
Here $\Omega_{\rm R}, \Omega_{\rm M}, \Omega_{\rm K}$ and $\Omega_{\Lambda}$
represent the fraction of the critical energy density ($3H^2c^2/8\pi G$)
in the form of radiation, matter, curvature and vacuum, respectively.
Each density component has an associated pressure, which is what makes
each evolve differently, with energy conservation leading to
$\rho\propto(1+z)^{3(1+w)}$.  The `equation of state' parameter
$w\equiv p/\rho c^2$, is thus $\frac{1}{3}$, 0, $-\frac{1}{3}$ and $-1$ for
radiation, matter, curvature and vacuum, respectively.

One of the motivations for exploring more general cosmological components
was the simple observation that there's something missing from the Friedmann
equation --
in other words, cosmologists asked themselves: `How come there's
no $(1+z)^1$ term?'  As we shall propose below, this term is not really
`missing', it is just `dark'.

This realization, coupled with the fact that $w\,{\leq}\,-\frac{1}{3}$ is
required for acceleration, led to the generalization of the cosmological
constant
to the so-called `quintessence' concept, i.e.~a component which is bracketed
by the properties of curvature and vacuum (and we can regard
$w\,{=}\,-\frac{1}{3}$
as simply another fluid, even within a flat Universe).

Measurements of the temperature and blackbody spectrum of the Cosmic
Microwave Background show that the photon energy density is very small in
today's Universe \cite{microwaves}.
We also know that the photon's less spin-challenged cousin, the graviton,
makes a negligible contribution.  Hence we can drop the $\Omega_{\rm R}$
component in the Friedmann equation.  The same would be true of any putative
new fluid with $w\,{=}\,\frac{2}{3}$, etc., since these would decay even more
rapidly as the Universe expands.

However, at the other end of the Friedmann equation we have components
which actually {\it increase\/} as the Universe expands, so-called
`phantom' DE \cite{phantom}.  This allows us to explore
components with $w\,{\le}\, -1$, which will make the Universe end with what has
been called `the Big R.I.P.'

Since there are many possibilities for dark energy, cosmologists often
use historical or philosophical principles to navigate among the choices.
One particularly useful idea was espoused by Walter of Ockham \cite{laser}
in the 14th century, when he said `{\it entia sunt
multiplicanda praeter necessitatem\/}', which is loosely translated as
`you can't get enough of a good thing'.  In other words, if dark energy is
so appealing, then let's have more of it!

One of the so-called `problems' with the Standard Model of Cosmology
\cite{SMC} is that there are several apparent coincidences with no
obvious physical explanation \cite{conspiracies}.  However, in this paper
we wish to stress the inverse of this, namely that the {\it most natural\/}
solution is that there are multiple forms of DE \cite{essence}.

We suggest here that the most reasonable cosmological model should have
each of the major constituents contributing equally to the cosmic energy
budget of a spatially flat Universe.  In other words we expect
$\Omega_{\rm D1}=\Omega_{\rm D2}=\Omega_{\rm D3}=\ldots = \frac{1}{N}$,
with $N$ being the total number of kinds of Darkness \cite{LetItBe}.

The model with the optimal level of naturalness has 6 Dark flavours, each
contributing $\frac{1}{6}$ to the total energy density.  Thus, in additional
to the common-or-garden Dark Matter, we also have
components which scale as $(1+z)^2$, $(1+z)^1$, $(1+z)^0$, $(1+z)^{-1}$ and
$(1+z)^{-2}$, and the individual values of $w$ are quantized in units of
$\frac{1}{3}$ \cite{quarks}.

This combination yields a current equation of state for the DE
(i.e.~non-material Darkness)
which is precisely $-1$, just as in the more traditional model.  The value
of $\Omega_{\rm DM}$ in the natural model is consistent with current Supernovae
and other cosmological data \cite{others}.

However, the model has a quite different evolution of $w(z)$.  Hence
even if it turns out to be difficult to precisely measure the value of
$\dot w$ in the past, all we have to do is wait for it to change in the
future!  In fact, since the naturalness of this model is only transient,
then there is a very strong prediction: sentient observers and the
cosmologists who exist alongside them, are fated not to last into future
eras when the model loses its naturalness.  Some theoretical constructs
are sometimes criticized for a perceived lack of predictability -- however,
no such criticism can be levelled against the Natural Dark Energy Hypothesis,
since it predicts nothing less than the end of civilization as we know it.

Since our hypothesis elegantly avoids any of the so-called
`coincidence' problems, we do not need to rely on any Anthropic reasoning to
explain the near-equality of the $\Omega$s.  However, there is nevertheless
one nagging issue, and that is: why is $\Omega_{\rm R}$ so very much smaller
than the other components?  We offer 4 possible non-Anthropic
resolutions of this `photon anomaly':
(1) the radiation is mainly composed of photons, which are
particles of light, and hence not bound by the rules which fix the behaviour
of the Dark sector \cite{information};
(2) perhaps one could invoke the `tired light'
hypothesis, in which the photons are partially absorbed by all the darkness
which they have to travel through \cite{light};
(3) the dark part of the radiation could
in fact {\it be\/} one of the Dark Energy components; or (4) {\it if\/} the
radiation were now a significant part of the cosmic energy budget, then the
Universe would be a very much hotter and more hostile place, and hence we
might not exist to observe it.

In this natural DE picture the different components presumably come out
of a complete fundamental theory, and hence it is also reasonable to expect
that the components could be coupled.  Exploration of {\it interacting\/}
parts of the Dark Side may help resolve other astrophysical puzzles.
As well as solving the Dark Matter and Dark Energy enigmas, we
have every reason to believe that our model will be just as successful
at solving the mysteries of the Cuspy Halo Problem,
Gamma-Ray Bursts, Ultra-High Energy Cosmic Rays, Baryon Asymmetry,
Primordial Magnetic Fields, the low CMB quadrupole, etc. \cite{further}



\smallskip



\baselineskip=1.6pt

\begin{thebibliography}{9}
\bibitem{legal} Citations removed following legal advice.
\bibitem{DE} E.g.~Albrecht A., et al., 2006, Report of the Dark Energy Task
Force, astro-ph:0609.591.
\bibitem{why120} If this was written instead as $66^{66}$ then it would appear
less mysterious, since there would be only one number to explain, rather
than 2.
\bibitem{crazy} For example: Einstein A., 1917, Sitzungsber. Preuss. Akad.
Wiss., 142; Weinberg S., 1987, PRL, 59, 2607;
Battye R.A. \& Moss A., 2007, astro-ph:0703.744.
\bibitem{microwaves} Which is lucky, since otherwise we would all be
cooked.
\bibitem{phantom} Caldwell R., 2002, Phys. Lett., B545, 23; Lucas G., et al.,
1999, SW, Ep.$\,1$.
\bibitem{laser} His older brother William also said some other things, but
those are not relevant here.  This principle, that if something has strong
merit then you want to make it as powerful as possible, is sometimes called
`Occam's Laser'.
\bibitem{SMC} For a particularly serious overview, see
Scott D., 2006, Can. J. Phys., 84, 419, astro-ph:0510.731.
\bibitem{conspiracies} Scott D. \& Frolop A., 2006, astro-ph:0604.011.
\bibitem{essence} We call these sextessence, septessence, octessence etc.
The ninth component would be nonessence, which clearly does not exist.
\bibitem{LetItBe} This may seem alarming to people who are afraid of the
Dark.  However, we {\it already\/} know that
the Universe is entering a period when the Dark Sector comes to dominate.
This is our Hour of Darkness, so we should just let it be.
\bibitem{quarks} Suggesting a connection with quark charges.
\bibitem{others} Of course there are several other possibilities, which,
although having slightly less naturalness, nevertheless may turn out to have
to be considered because they are a better fit to the data.  Other examples
have: 5 components with $\Omega_{{\rm D}i}=0.2$ and $w\,{=}\,-\frac{5}{6}$
or $-1$; or 4 components with $\Omega_{{\rm D}i}=0.25$, e.g.~with the Dark
Energy having terms of the form $(1+z)^{3/2}$, $(1+z)^0$ and $(1+z)^{-3/2}$,
so that $w\,{=}\,-1$ still.
\bibitem{information} There may be a form of Dark Entropy which dominates
over the normal radiation, perhaps stopping us from reading the full
information content of the CMB; see Scott D. \& Zibin J.P., astro-ph:0511.135.
\bibitem{light} Terry Pratchett wrote `Light thinks it travels faster than
anything but it is wrong.  No matter how fast it travels, it finds the
darkness has always got there first, and is waiting for it.'
\bibitem{further} Further suggestions should be sent to Dr.~Frolop.
\end{thebibliography}
\end{document}